\documentclass[11pt]{article}

\usepackage{fullpage}
\usepackage{amsmath}
\usepackage{amsthm}
\usepackage{amssymb}
\usepackage{authblk}

\newtheorem{theorem}{Theorem}[section]
\newtheorem{lemma}[theorem]{Lemma}

\newtheorem{definition}[theorem]{Definition}

\newtheorem{corollary}[theorem]{Corollary}

\newtheorem{proposition}[theorem]{Proposition}

\usepackage{wrapfig}

\newcommand{\cref}[1]{\texttt{#1}}

\usepackage{tikz-cd}

\usepackage{graphicx}
\graphicspath{{image/}}

\usepackage{hyperref}
\hypersetup{
    colorlinks=true,
    linkcolor=violet,
    filecolor=violet,
    urlcolor=violet,
    citecolor =violet,
    pdfpagemode=FullScreen,
    }
\newcommand{\e}{\varepsilon}

\usepackage[backend = biber, url=false,sorting=nyt,style=numeric-comp,maxbibnames = 99]{biblatex}
\newcommand{\inv}{^{-1}}
\newcommand{\R}{\mathbb{R}}
\newcommand{\RR}{\mathcal{R}}
\newcommand{\Reeb}{\mathbf{Reeb}}
\newcommand{\PP}{\mathcal{P}}

\newcommand{\Pers}{\mathbf{Pers}}

\newcommand{\Ord}{\mathrm{Ord}}
\newcommand{\rel}{\mathrm{Rel}}
\newcommand{\Rel}{\mathrm{Rel}}
\newcommand{\Ext}{\mathrm{Ext}}
\newcommand{\Id}{\mathrm{Id}}

\renewcommand{\phi}{\varphi}

\begin{document}
\title{Realizable piecewise linear paths of persistence diagrams\\ with Reeb graphs}
\author[1]{Rehab Alharbi}
\author[2]{Erin Wolf Chambers}
\author[2]{Elizabeth Munch}
\affil[1]{Dept of Mathematics \& Statistics, Saint Louis University}
\affil[2]{Dept of Computer Science, Saint Louis University}
\affil[3]{Dept of Computational Mathematics, Science \& Engineering, Dept of Mathematics, Michigan State University}
\date{}

\maketitle

\begin{abstract}

Reeb graphs are widely used in a range of fields for the purposes of analyzing and comparing complex spaces via a simpler combinatorial object.
Further, they are closely related to extended persistence diagrams, which largely but not completely encode the information of the Reeb graph.
In this paper, we investigate the effect on the persistence diagram of a particular continuous operation on Reeb graphs; namely the (truncated) smoothing operation.
This construction arises in the context of the Reeb graph interleaving distance, but separately from that viewpoint provides a simplification of the Reeb graph which continuously shrinks small loops.
We then use this characterization to initiate the study of inverse problems for Reeb graphs using smoothing by showing which paths in persistence diagram space (commonly known as vineyards) can be realized by a path in the space of Reeb graphs via these simple operations.
This allows us to solve the inverse problem on a certain family of piecewise linear vineyards when fixing an initial Reeb graph.

\end{abstract}

\section{Introduction}

Reeb graphs have become an important tool in topological data analysis for the purpose of visualizing  continuous functions on complex spaces, as they yield a simplified discrete structure.
Originally developed in relation to Morse theory~\cite{Reeb1946}, these objects are used extensively for shape comparison, constructing skeletons of data sets, surface simplification, and visualization; for more details on these and more applications, we refer to recent surveys on the topic~\cite{Biasotti2008,yan2021scalar}.
While some information is lost in the construction of the Reeb graph, such simplified structures allow for more efficient methods to analyze and compare data sets.

\begin{figure}%
\centering

    \includegraphics[width=0.3\textwidth]{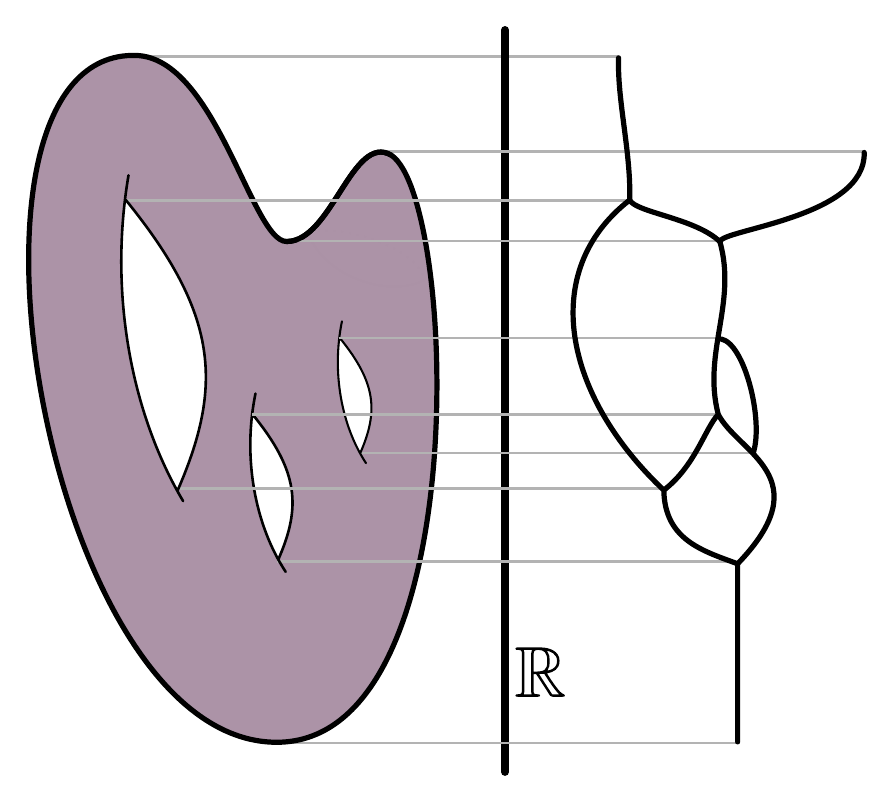}
    \caption{Reeb graph (right) for a given space with real valued function (left) defined by height.}
  \label{rspace244}
\end{figure}

More precisely, given a topological space $M$ and  a real valued function $f: M \to \mathbb{R}$, the pair $(M,f)$ is known as an $\mathbb{R}$-space; e.g.~see Figure~\ref{rspace244}.
The Reeb graph of $f$ is then obtained by  collapsing each connected component in a level set into single point, and collecting the points together using the quotient topology.  The result is a 1-dimensional stratified space (i.e.~a graph) along with an induced real-valued function; we will refer to this pair as $(X,f)$.

Given the many algorithms available to compute these objects efficiently~\cite{Doraiswamy2009,Parsa2012,Harvey2010,Gueunet2019}, the Reeb graph is a practical, simplified structure which can be used for tasks ranging from simplification to visualization.
Thus, there is a practical need for ways to compare and analyze Reeb graphs.
Many possible options have been studied recently
\cite{DiFabio2016,Bauer2020,deSilva2016,chambers2020family,Bauer2014,Carriere2017,Cohen-Steiner2009a,Bauer2015b,Bauer2014,deSilva2016,Bauer2020}.
In this work, we focus on recent work \cite{deSilva2016,chambers2020family} which introduces the concept of smoothing a Reeb graph as a byproduct of the interleaving distance.
This smoothing operation generates a new Reeb graph $S_\e(X,f)$ for every $\e \geq 0$ which simplifies the topological structure of the graph.
In particular, it continuously removes small loops, which are often viewed as noise in the input data.

Despite being defined via category theory, the resulting construction can be viewed from a completely combinatorial viewpoint.
An algorithm for constructing the smoothed Reeb graph was provided in~\cite{deSilva2016}, and has been extended to the more recently introduced truncated smoothing functor as well~\cite{chambers2020family}.
Further, it has been shown that given a Reeb graph $(X,f)$ with critical points $S = \{a_i\}$,  the smoothed Reeb graph $S_\e(X,f)$  has critical set contained in $S_\e = (S-\e) \cup (S+\e) = \{ a-\e, a + \e \text{ } | \text{ } a \in S \}$~\cite{deSilva2016}.
However, no exact combinatorial characterization of the graph changes has been proven in the literature prior to this paper.

We note that $\R$-space data can also be studied via its persistent homology, an algebraic method for measuring topological features of shapes and functions~\cite{Oudot2017a,Dey2021}.
In fact, that there is a close relationship between critical values of the Reeb graph and points in the extended persistence diagram \cite{Agarwal2006,CohenSteiner2006}, which will be a key component of our work.
Because of this relationship, one might expect there to be an available inverse map; i.e.~given an extended persistence diagram, can we reconstruct the Reeb graph uniquely?
The answer, in general, is no since many different Reeb graphs can have the same persistence diagram.
Despite difficulties, consideration of these inverse problems are increasingly of interest in topological data analysis as a whole~\cite{curry2019left,Oudot2020,Curry2021}, as they can provide insight into how much information is lost in the process of computing such a topological signature, be it a persistence diagram or a Reeb graph.
However, they are notoriously difficult, as key information is inevitably lost when going from a space to such a signature.

\subsection{Outline of our results}
The main contribution of this paper is to initiate the study of inverse problems on vineyards realized by Reeb graphs, using the smoothing and truncated smoothing functors to determine when such an inverse can be determined.
After introducing necessary background and notation in Section~\ref{sec:background},  in Section~\ref{sec:geometricanalysis} we explicitly enumerate all changes to a Reeb graph under Reeb graph smoothing, and
then generalize this framework for the more recently developed truncated smoothing operation.
In Section~\ref{sec:inverseproblems}, we use this characterization to solve the inverse problem in a restricted setting for Reeb graphs using smoothing and truncated smoothing, by showing that certain paths in persistence diagram space (commonly known as vineyards \cite{Cohen-Steiner2006}) can be realized by a path in the space of Reeb graphs.
Namely, using our characterization of how truncated smoothing affects the combinatorial structure of the Reeb graph, we are able to determine sufficient restrictions on a time varying set of persistence diagrams, so that as long as an initial Reeb graph is specified, we can  determine a set of time varying Reeb graphs which realize the vineyard.   We conclude in Section~\ref{sec:conclusion} by discussing several possible future directions motivated by our work.

\section{Background and definitions}
\label{sec:background}

\subsection{Reeb graphs}

Let X be a topological space, and let $f: X \to$ $\mathbb{R}$ be continuous real-valued function.
The pair $(X,f)$ is  referred to as an $\mathbb{R}$-space (or a scalar field, depending upon the setting).
The level set of $f$ at $a$ is the set $f^{-1}(a) =\{x \in X \mid f(x) = a\}$.
We define an equivalence relation $\sim_f$ on $X$ by $x \sim y$ if and only if $f(x)=f(y)=a$ and $x$ and $y$ are in the path same connected component of the levelset $f^{-1}(a)$.
The Reeb graph $\RR(X,f)$ is the quotient space $X/_\sim$, with an induced function inherited from the $\mathbb{R}$-space given by $\overline{f}([x]) = f(x)$.  (We will generally abuse notation slightly and call both functions $f$.)

For the purposes of our work, we will often divorce the idea of a Reeb graph from the $\R$-space it came from.
In particular, given reasonable assumptions on the input $\R$-space, the Reeb graph is indeed a finite graph, so we will assume that our graphs have this property.
We will assume further that all Reeb graphs are \emph{constructible},  defined as follows.

\begin{definition}
An $\R$-space is constructible if it is homeomorphic to one constructed in the following manner.
We are given a finite set of critical points $a_1, \ldots, a_n$,  and a collection of spaces $\{V_i\}_{i=1}^n$ and $\{ E_i\}_{i=1}^{n-1}$.
Further, we specify left attaching maps $\ell_i:E_i \to V_i$ for $i = 1,\cdots n$ and right attaching maps $r_i:E_i \to V_{i+1}$.
We then define $X$ to be the quotient space
\begin{equation*}
\coprod_{i=1}^n
    (V_i \times \{a_i\}
\coprod_{i=1}^{n-1}
    (E_i \times [a_i \times a_{i+1}]
\end{equation*}
with respect to the identifications $(\ell_i(e), a_i) \sim (e,a_i)$ and $(r_i(e),a_{i+1}) \sim (e,a_{i+1})$.
We define the function $f$ to be the projection on the second factor.

A constructible $\R$-space is a Reeb graph if all the $V_i$ and $E_i$ are discrete, finite sets of points.
\end{definition}

In particular, a Reeb graph can be encoded by the combinatorial data of a finite graph $X$, and a function defined on the vertices $f:V \to \R$.
This can be extended to the edges linearly, and in this case, we require that no adjacent vertices have the same function value.
We call the number of edges incident to $v$ that have higher values of $f$ its {\it{up-degree}}, and define the term {\it{down-degree}} symmetrically.
We always assume that a vertex of both up- and down-degree 1 is replaced with the relevant edge; this way, every vertex in the Reeb graph is a critical point.
A vertex $v$ is local maximum (local minimum) if it has up-degree $0$ (down-degree $0$).
Likewise, a vertex is an up-split (down-split) if it has up-degree (down-degree) strictly larger than $1$.
Note that a vertex can be both an up-split and a down-split; or an up-split and a local minimum, etc., although this does not happen for a generic Morse function on a constructible $\mathbb{R}$-space

In this paper, we will primarily restrict our attention to Reeb graphs with fairly strong genericity assumptions in order to simplify proofs.
Our notion of genericity will involve two distinct criteria.
The first is when we have more than one critical point at a given function value; if no such pair exists, we say the graph is \textit{function-generic}.
The second possible violation is if the vertex is of a type not seen in the case of Morse functions on manifolds.
The four kinds of vertices which can appear in this case, with their (down-, up-) degrees specified, are local minima (0,1);  up-forks (1,2);  down-forks (2,1); and local maxima (1,0).
If a Reeb graph has only vertices of these four types, it is called \textit{Morse-generic}.
If a Reeb graph is both function- and Morse-generic, we simply call it \textit{generic}.

In the case of Morse-generic Reeb graphs, we have a strong characterizations of the critical points.
We say a down-fork $v$ is an \emph{ordinary down fork} if the two lower branches of $v$ are contained in different connected components of of the open sublevel set $\RR(G)_{<a}:= f\inv(-\infty,a)$.
Otherwise, we say $v$ is an \emph{essential down fork}.
The ordinary and essential up-forks are defined in the same way, using the open super-level set $\RR(G)_{>a}:= f\inv(a, \infty)$.

\subsection{Smoothing and truncated smoothing}

We now turn our attention to the (geometric) definition of smoothing given in \cite{deSilva2016}, and the truncated smoothing given in \cite{chambers2020family}.
Both smoothing and truncated smoothing were studied in the context of comparing two Reeb graphs, where the focus was on defining distances between the graphs.
Both allow for the definition of an interleaving distance with desirable theoretical properties~\cite{deSilva2016,chambers2020family}.  While these distances partially motivate our study, we do not directly use these interleaving distances in our work, but rather focus on the two operations themselves.

\begin{figure}
    \centering
    \includegraphics[width = .5\textwidth]{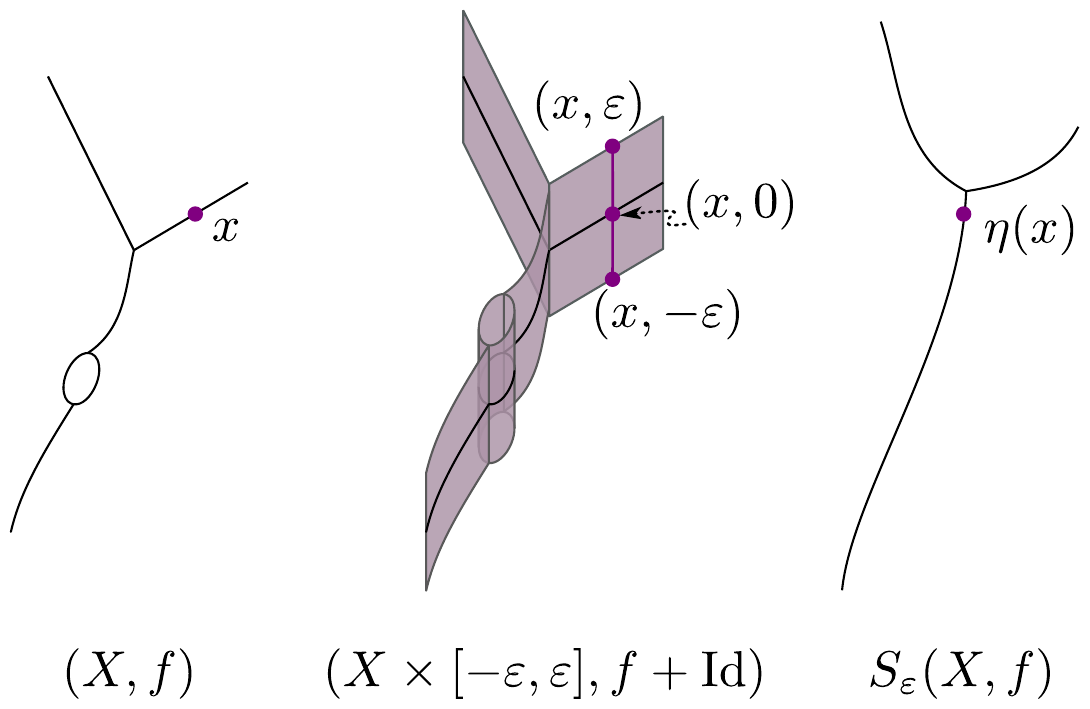}
    \caption{An example of the thickening and smoothing procedure. Given a Reeb graph (left) with function implied by height, we cross the graph with an interval $[-\e,\e]$, drawn so that the induced function $f_\e$ is still visualized by height. Then the smoothed Reeb graph $S_\e(X,f)$ is the Reeb graph of this $\R$-space. }
    \label{fig:ConstructingSmoothing}
\end{figure}

Let $(X,f)$ be a Reeb graph and let $\e \geq 0$.
Define $(f+Id): X\times [-\e, \e] \to \mathbb{R} $ by $(x,t)\mapsto f(x) + t$.
We define the \emph{$\e$-smoothing} $S_\e(X,f)$ to be the Reeb graph of $(X \times [-\e,\e], f+Id)$; we denote the corresponding quotient map by $q: (X,f) \times [-\e,\e] \to S_\e(X,f)$.
Further, we denote the induced map $q \circ (f+\Id) = \eta:(X,f) \to S_\e(X,f)$; i.e.~$\eta$ is defined so that the diagram
\[
\begin{tikzcd}
 & (X,f) \times [-\e,\e] \arrow{dr}{q} \\
(X,f) \arrow{ur}{i} \arrow{rr}{\eta} && S_\e(X,f)
\end{tikzcd}
\]
commutes.
See Fig.~\ref{fig:ConstructingSmoothing} for an example.

It is worth noting that smoothing is a functor, with further structure that we will not utilize in this paper; see~\cite{deSilva2016,deSilva2018} for details.
We will focus on the combinatorial properties of smoothing and the $\eta$ map in our work, and will  characterize how a Reeb graph changes under $\eta$ as $\e$ varies.
We begin with the following lemma that shows $\eta$ does not change the connected components; this result is implicitly referenced in prior work~\cite{deSilva2016} but never explicitly proven, so we include a proof here for completeness.

\begin{lemma}
The induced map  $\pi_0[\eta]$ gives an isomorphism
$\pi_0 (X,f) \cong \pi_0 (S_{\e}(X,f)).$
\end{lemma}

\begin{proof}
Note that because $\eta = q \circ i$, functoriality says that we need only show that $\pi_0[q]$ and $\pi_0[i]$ are isomorphisms as $\pi_0[\eta] = \pi_0[q] \circ \pi_0[i]$.

From Theorem 23.6 of Munkres~\cite{munkres2000topology}, a finite Cartesian product of connected spaces is connected.
So for each connected component $A \subseteq G$, $A \times [-\e,\e]$ is connected, and thus $\pi_0[i]: \pi_0(X,f) \to \pi_0((X,f) \times [-\e,\e])$ is an isomorphism.

The map $\pi_0[q]: \pi_0(X \times [-\e,\e]) \to \pi_0(S_\e(X,f))$
is surjective because by definition of a quotient map, $q$ is surjective.
We next show $\pi_0[q]: \pi_0(X \times [-\e,\e]) \to \pi_0(S_\e(X,f))$ is  injective.
By Munkres Exercise 23.11~\cite{munkres2000topology}, since our quotient map $q$  has $q \inv(y)$ connected, then for each connected component $A$ of $S_\e(X,f)$, $q \inv( A) $ is connected.
This implies that two connected components of $X \times [-\e,\e]$ cannot map to the same connected component of $S_\e(X,f)$ under $q$ without a contradiction.
Thus $\pi_0[q]$ is injective, finishing the proof.
\end{proof}

 \begin{figure}%
  \includegraphics[width=\textwidth]{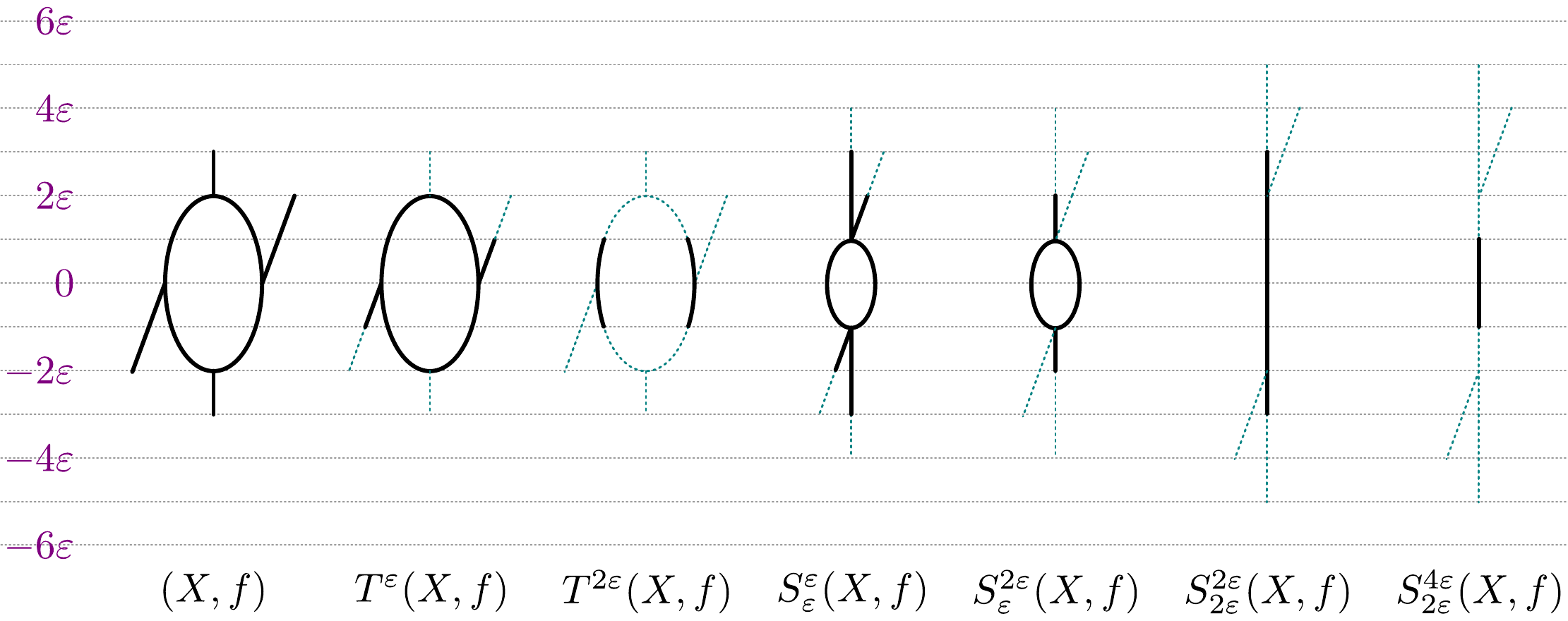}
    \centering
  \caption{ Examples of smoothing and truncating a given (left most)  Reeb graph.  The sets $U_\tau(X,f)$ and $D_\tau(X,f)$ from $S_\e(X,f)$ are indicated as dotted lines in the graphs to the right, which are removed in the truncated graph. Note that the notation $S_\e^\tau = T^\tau S_\e$ denotes truncation after smoothing.}
  \label{fig:graphandSmoothing}
\end{figure}

Truncated smoothing is a more recently developed variation of the smoothing functor, which intuitively smooths and then ``chops off" tails in the smoothed graph~\cite{chambers2020family}.
More formally,  define a path in the Reeb graph $(X,f)$ to be a map $\gamma:[0,1]\to X$.
This path is monotone increasing (resp.~decreasing) if $f(\gamma(t)) \leq f(\gamma(t')) $ (resp.~$f(\gamma(t)) \geq f(\gamma(t')) $) for all $t \leq t'$.
We also call these up- and down-paths, respectively.
The height of a monotone path $\gamma$ is $|f(\gamma(0)) - f(\gamma(1))|$.
Let $U_\tau (X,f)$ be the set of points of $X$ that do not have a height $\tau$ up-path and let $D_\tau$ be the set of points of $X$ that do not have a height $\tau$ down-path.
 The truncation of Reeb graph $(X,f)$ is defined by
$$T^{\tau} (X,f)= (X,f) \setminus (U_\tau  \cup D_\tau)$$
so that we keep only the subgraph of $(X,f)$ that consists of the points that have both up-path and down-path of height $\tau$.
The truncated smoothing of Reeb graph  $(X,f)$ is defined to be
\begin{equation}
\label{eq:truncSmoothingDefn}
S^{\tau}_{\e}(X,f) = T^\tau S_\e (X,f).
\end{equation}
See Fig.~\ref{fig:graphandSmoothing} for examples.

The truncated smoothing operation inherits many of the useful properties of regular smoothing.
First, $S_\e^\tau$ is a functor and we have a map $\eta: (X,f) \to S_{\e}^{\tau}(X,f)$ for any $0 \leq \tau \leq \e$.
Note that we abuse notation and write $\eta$ since this map is a restriction of the map $\eta:(X,f) \to S_{\e}(X,f)$.\footnote{This map is denoted $\rho$ in \cite{chambers2020family}.}.
We note also that $\pi_0(S_\e(X,f)) \simeq S_\e^\tau(X,f)$, since it is shown in \cite{chambers2020family} that if $0\leq \tau \leq 2\e$ and $(X,f)$ is connected then $S_\e^\tau(X,f)$ is connected.

\subsection{Persistent Homology}

In this section, we give a brief introduction to extended persistent homology.
We assume that the reader is familiar with standard homology and direct the reader to \cite{munkres2000topology,Hatcher} for further details, as well as to \cite{Oudot2015} for more details on (extended) persistent homology.

First, given any filtration of topological spaces
\begin{equation*}
    X_1 \subseteq X_2 \subseteq \cdots \subseteq X_n
\end{equation*}
functoriality gives a sequence of vector spaces and linear transformations
\begin{equation*}
    H_p(X_1) \to H_p(X_2) \to \cdots \to H_p(X_n).
\end{equation*}
For the expert, we note that we assume that our homology is computed over a field $\mathsf{k}$ so that the resulting spaces are indeed vector spaces.
We define a \emph{persistence module} to be a sequence of vector spaces and linear transformations of the form
\begin{equation*}
 \mathbb{V}:= \{ V_1 \to V_2 \to \cdots \to V_n\}.
\end{equation*}
An interval module $\mathbb{I} = \{ I_1 \to \cdots \to I_n\}$ is a persistence module where $I_i=0$ is the trivial vector space for $i \not \in [a,b)$, $I_i =\mathsf{k}$ is the 1-dimensional vector space otherwise, and the linear maps are isomorphisms when possible and 0 otherwise.
By~\cite{Zomorodian2004} (see also~\cite[Thm.~1.9]{Oudot2015}), any pointwise finite dimensional persistence module can be written as a direct sum of interval modules
\begin{equation*}
    \mathbb{V}= \bigoplus_{(a,b) \in \mathcal{B}} \mathbb{I}_{[a,b)}.
\end{equation*}
This representation is unique up to isomorphism, so we can draw the information as a persistence diagram: a multiset of points $\{ (a,b) \in \mathcal{B}\}$ in the upper half plane.
For technical reasons related to computing distances, we also consider every diagram to have trivial points on the diagonal $\Delta = \{ (x,x) \mid x \in \R\}$.
We often refer to a point $(a,b)$ in the persistence diagram as representing a class \textit{born} at $a$ and \textit{dying} entering $b$.

Given a constructible $\R$-space $(X,f)$ with critical set $a_1,\cdots,a_n$, the (standard) persistence diagram is defined to be the above decomposition for the filtration
\begin{equation*}
    H_d(X_{a_1}) \to H_d(X_{a_2}) \to \cdots \to H_d(X_{a_n})
\end{equation*}
for the (closed) sublevel sets $X_a:= f\inv(-\infty,a]$.
Here, the maps on homology are induced by inclusion $X_a \subseteq X_b$ for $a \leq b$.
When necessary, we specify the dimension $d$ of homology used by calling this a $d$-dimensional diagram.

In the case of a Reeb graph $(X,f)$, the only interesting portions of the standard persistence diagrams are that local minima of the form $f(v) = a$ give rise to points in the 0-dimensional diagram with an infinite class $(a,\infty)$ for the global minimum of  each connected component; and essential down forks of the form $f(v) = a$ give rise to points $(a,\infty)$ in the 1-dimensional persistence diagrams (i.e., the loop is born at the top).
Clearly, there is more obvious homological structure in these Reeb graphs, so we instead turn to the \emph{extended persistence diagram}~\cite{Agarwal2006,CohenSteiner2008}, defined as follows.

We extend our ordinary persistence module by using the relative homology of the super level sets.
Write  $X^a = f\inv[a,\infty)$ for the super-level sets.
Note that $X = X_{a_n}$; $X^{a_n}$ is a discrete set consisting of the points with function value at the global maxima; and $H_d(X,X^{a_1}) = 0$.
Further, $H_d(X_{a_n}) = H_d(X) = H_d(X,\emptyset)$, so we have a map $H_d(X) \to H_d(X,X^{a_n})$.
Thus, we build the extended persistence module
\begin{equation*}
0 \to H_d(X_{a_1})
\to \dots \rightarrow
H_d(X_{a_n})
\rightarrow
H_d(X,X^{a_n})
\rightarrow \dots \rightarrow
H_d (X,X^{a_1})=0.
\end{equation*}
We can decompose the resulting persistence module as before, but represent the intervals slightly differently by associating them to the indices $a_i$ and $a_j$ for where the interval starts and ends, while encoding whether these endpoints happen in the first or second half of the persistence module by putting them in a different sub-diagram.
In a Reeb graph, the resulting \emph{extended persistence diagram} can be decomposed into four sub-diagrams \cite{Agarwal2006}.

First, there are classes which are born and die in the first half of the module; we represent these by so-called ordinary persistence points since they correspond to finite-lifetime points which would show up in the traditional persistence diagram.
In the case of a Reeb graph, these only appear in dimension 0, and the ordinary sub-diagram, $\Ord_0$, contains a point $(a_i,a_j)$ with $i \le j$ for each non-infinite bar in the traditional persistence diagram.

The second type of points come from bars in the persistence module which are entirely contained in the second half of the diagram.
We represent an interval lasting from $H_d(X,X^{a_j}) \to H_d(X,X^{a_i})$ at $(a_j,a_i)$ in the extended persistence diagram, noting that $j\geq i$ so these points are always below the diagonal.
In the case of Reeb graphs, these kinds of points can only appear in dimension 1, so we call this the relative sub-diagram, $\Rel_1$.

The last kind of intervals that appear in the persistence module are those that begin at $H_d(X_{a_i})$ and end at $H_d(X,X^{a_j})$; we then include a point in the extended diagram at $(a_i,a_j)$.
In this case, we can have either $i \leq j$ or $j \leq i$ so points can be both above and below the diagonal in the diagram.
In the case of Reeb graphs, these kind of intervals can appear in both dimensions 0 and 1, so we have the extended sub diagrams $\Ext_0$ and $\Ext_1$.

\begin{figure}
    \centering
    \includegraphics[width = .9\textwidth]{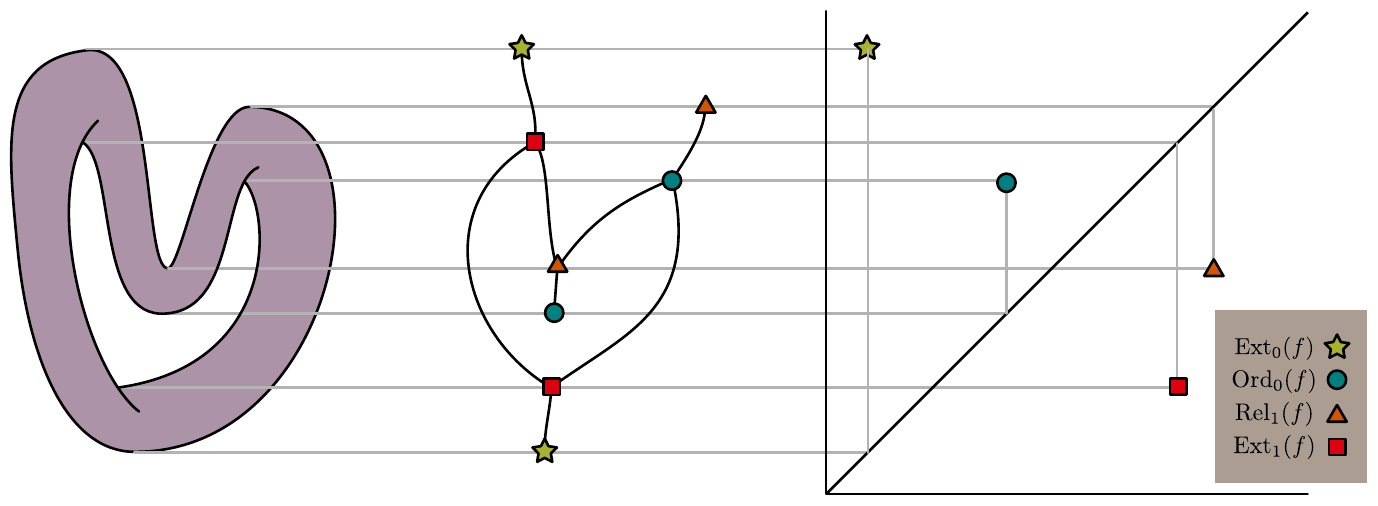}
    \caption{An example of a Reeb graph and its extended persistence diagram. }
    \label{fig:ExtendedReebExample}
\end{figure}

In the case of a Morse-generic Reeb graph, we have a complete pairing of the vertices with respect to the extended persistence diagram.
Each point in the $\Ext_0$ diagram corresponds to a connected component of the Reeb graph, born at the global minimum value in the first half of the module, and dying at the global max value in the second half of the module.
Each point in the $\Ext_1$ diagram corresponds to a loop in the Reeb graph, born in the first half of the persistence module at the highest function value vertex in the loop, and dying in the second half of the persistence module at the lowest function value vertex.
Note that these are always essential up- and down-forks.
Points in $\Ord_0$ and $\Rel_1$ correspond to pairs of vertices: local minima with ordinary down forks in the first case and local maxima with ordinary up forks in the second case.
See Figure~\ref{fig:ExtendedReebExample} for an illustration where symbols corresponds to the pairing induced by the extended persistence diagram.

When we lose the Morse-generic assumption, this pairing still is present but is a bit more subtle.
In particular, it is possible to have a vertex in the Reeb graph which contributes to more than one type of persistence point.
More precisely, an up-split with $d > 1$ up edges will contribute to more than one different persistence points; down-splits behave analogously.
For this reason we will state our main theorem in the context of generic Reeb graphs, although with some bookkeeping it can be modified to the case of non-generic Reeb graphs.

\subsection{Bottleneck distance}

The bottleneck distance first arose in the context of persistent homology, where it was used to assess stability in persistence diagrams~\cite{Steiner2005}.
Intuitively, this distance between diagrams considers all pairings of points of the same dimension and type in the two diagrams and calculates the maximum distance among all matched pairs under the $L_\infty$ norm.
Note that points are also allowed to match to the diagonal to compensate for the potential of having a different number of off-diagonal points in the diagrams under consideration.
In a sense, we can think of this as overlaying the extended persistence diagrams of both graphs and then matching points either to a point of the same type, or to the diagonal.

\begin{definition}
Let $(X,f),(Y,g)$ be two Reeb graphs.
We define the \textbf{bottleneck distance} $d_b$ between their extended persistence diagrams $D(X,f)$ and $D(Y,g)$ as
\[
d_{b}(\emph{D}(X,f),\emph{D}(Y,g))
= \inf_{m} \sup_{x \in \emph{D}(X,f)}||x - m(x)||_{\infty},\]
where $m$ is a bijection between the multiset of points of $\emph{D}(X,f)$ and $\emph{D}(Y,g)$ (including points on the diagonal),
and the bijection must match points of the same type ($Ord_0(f)$ to $Ord_0(g)$, $Rel_1(f)$ to $Rel_1(g)$,  etc.)

\end{definition}

\begin{figure}
    \centering
    \includegraphics[width = .9\textwidth]{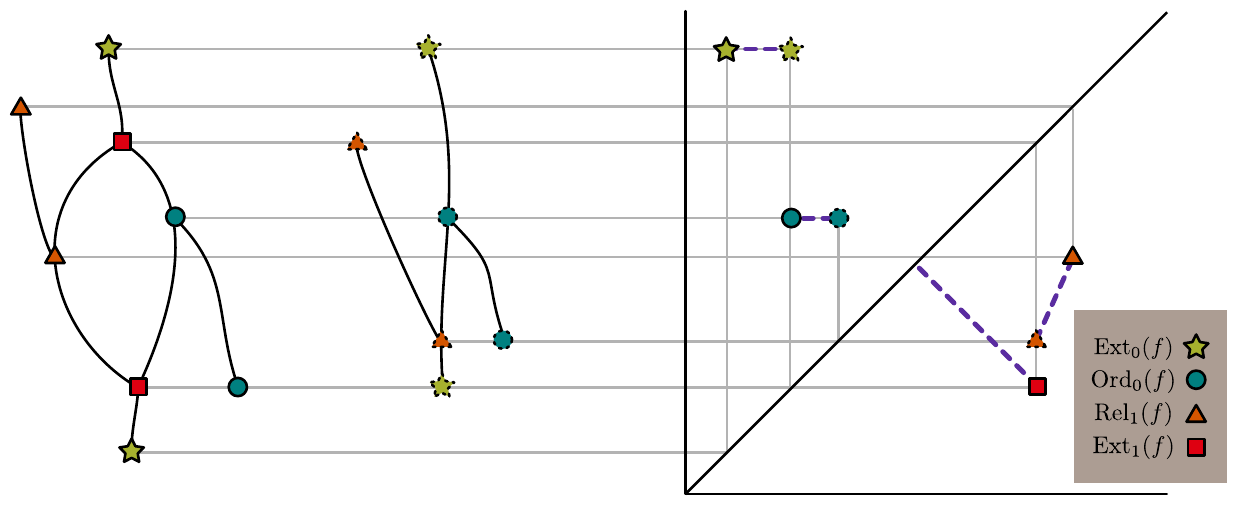}
    \caption{A visualization of the bottleneck distance between the extended persistence diagrams of two Reeb graphs. Note that the pairing from the bottleneck distance must match points of the same type, so even though the solid square and dashed triangle at the bottom right of the diagram are close, the bottleneck matching cannot pair them together. }
    \label{fig:bottleneck}
\end{figure}

See Fig.~\ref{fig:bottleneck} for a visual representation of the bottleneck matching between two persistence diagrams.

\section{Geometric analysis of smoothing}
\label{sec:geometricanalysis}

In this section, we analyze the smoothing operation more carefully and precisely characterize the behavior of the critical set during the smoothing based on which part of the extended persistence diagram it contributes to.
Geometrically, this implies that  smoothing eliminates cycles whose height is less than $2\e$, and that smoothing eventually results in the graph becoming a forest.  The proof of this theorem depends upon a more  detailed geometric classification of how smoothing affects down- and up-forks.

We first need the following lemma, proven in \cite{deSilva2016}, which specifies properties of the projection map works in smoothing:
\begin{lemma}\cite[Lemma 4.22]{deSilva2016} \label{lem:homotopy_equiv}
Let $p: X \times [-\e, \e] \rightarrow X$ be the projection map onto the first factor. Then the map $p$ restricts to a homotopy equivalence $f^{-1}_\e (I) \simeq f^{-1}(I^\e)$.
\end{lemma}

This lemma allows us to completely characterize vertices in the smoothed Reeb graph by looking at inverse images from our original Reeb graph rather than needing to work with the thickened version, $X \times [-\e,\e]$.
In particular, we can investigate $S_\e(X,f)$ at function value $b$ to characterize when there is a vertex at that value, and what its up- and down-degrees are.

For the sake of notation, denote by $[b]^\e = [b-\e,b+\e]$ and $(b)^\e = (b-\e,b+\e)$ the closed and open intervals of width $2\e$, respectively.
Taking the limit of Lem.~\ref{lem:homotopy_equiv} over intervals $(b)^\delta$ as $\delta \to 0$, we have that $\pi_0f_\e\inv(b) \cong \pi_0f\inv([b]^\e)$; this gives the points in $S_\e(X,f)$ at the function value $b$.
Because $(X,f)$ is constructible, for a small enough $\delta>0$ the points in $S_\e(X,f)$ immediately below are elements of $\pi_0f\inv([b-\delta]^\e)$, while the points immediately above are elements of $\pi_0f\inv([b+\delta]^\e)$.
Intuitively, the points at $b$ are regular if there is exactly one up and one down path emanating from them; deviating from this results in that point being a vertex.

To determine which elements of these sets correspond to a vertex, we can keep track of the attaching maps as follows.
First, note that by assumption,  $[b]^\e$ and $[b]^{\e + \delta}$ intersect the same collection of critical values of $(X,f)$, thus by constructibility, the map induced by inclusion $\pi_0f\inv([b]^\e) \to \pi_0f\inv([b]^{\e+\delta})$ is an isomorphism.
The point of passing to this larger interval is that now $[b-\delta]^\e$ and $[b+\delta]^\e$ are both contained in it, so we have the diagram
\begin{equation}
\label{eq:IntervalDgm}
    \begin{tikzcd}
    & \pi_0f\inv[b]^{\e+\delta}\\
    \pi_0f\inv[b-\delta]^{\e} \ar[ur, "\alpha"]
    && \pi_0f\inv[b+\delta]^{\e}\ar[ul, "\beta"]\\
    & \pi_0f\inv[b]^{\e}. \ar[uu, "\cong"]
    \end{tikzcd}
\end{equation}
In $S_\e(X,f)$, the number of points at $b$ is in the bottom set, the lower edges at left, and the upper edges at right.
So $\alpha$ and $\beta$ give attaching information for the lower and upper edges, respectively.
Further, there is a vertex at function value $b$ if $\alpha$ or $\beta$ (or both) are not isomorphisms.
See Fig.~\ref{fig:GraphsWithIntervals} for an example of these interval representations.
We will use this setup repeatedly in the next section to characterize vertices and attaching information to understand how the vertices move in the smoothed version of the Reeb graph.

\subsection{Smoothing Reeb Graphs}

Next, we consider how to completely determine the  combinatorial structure of $S_\e(X,f)$.  In fact, prior work considered this:  \cite[Corollary 4.25]{deSilva2016} claims that the set of critical points of $S_\e(X,f) = \{ a \pm \e \mid a \in S \}$, where $S$ is the set of critical points of the Reeb graph $(X,f)$.
In fact, there is some nuance to this issue, as $S_\e(X,f)$ is a subset of $\{ a \pm \e \}$, but (assuming we reduce vertices of degree 2 and keep only vertices of degree 3 or more as critical values) it is never equal to the set $\{ a \pm \e \}$.

In this paper, we work exclusively with generic Reeb graphs $(X,f)$ and $\e$ values which keep the graph $S_\e(X,f)$ generic as well.
However, even with a generic input Reeb graph, particular choices of $\e$ can result in non-generic $S_\e(X,f)$, as smoothing intuitively will move vertices to new function values where they have the potential to "bump into'' other vertices.
Our proof can be adapted to work for non-generic Reeb graphs and the attaching information described in the last section still completely determines the combinatorial changes; however, we limit our proof to only generic graphs in order to simplify the case analysis.

For a generic Reeb graph $(X,f)$, we completely characterize the critical set of $S_\e(X,f)$ as follows:

\begin{theorem}
\label{thm:bijection}
Let $(X,f)$ be a generic Reeb graph with  vertex set
$V(X) = \{ v_1, v_2, \ldots, v_n\}$.
We denote the critical set
$S = \{a_1 < a_2 < \ldots <  a_n\}$
and assume that the vertices are sorted so that $f(v_i) = a_i$.
Let $\e >0$ be a value such that
 $2\e \ne |a_i - a_j|$ for any $i,j$.
Let $W\subseteq V$ be the subset of vertices where each $w \in W$ contributes to a point in  $\Rel_1$ which has lifetime at most $2\e$.
Then there is a bijection
$$\Phi: V(X) \setminus W \to  V(S_\e(X)),$$
and $S_\e(X,f)$ is generic.
\end{theorem}

\begin{figure}[t]
    \centering
\includegraphics[width = .5\textwidth]{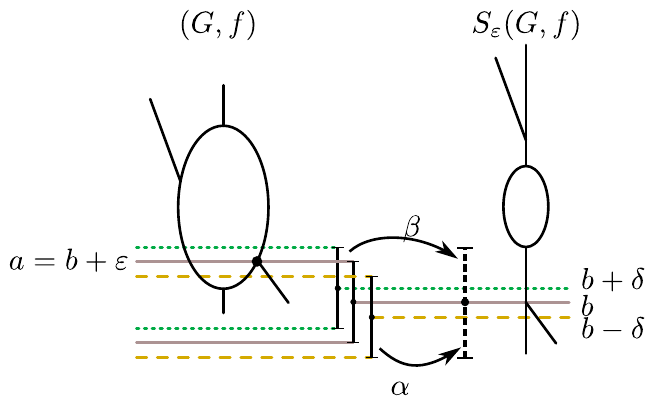}

 \caption{The intervals $[b]^\e$, $[b-\delta]^\e$, and $[b+\delta]^\e$ as well as the dashed interval $[b]^{\e+\delta}$ shown in between a Reeb graph (left) and its smoothing (right); the inverse images in $(X,f)$ are indicated by dotted lines, and attaching maps $\alpha$ and $\beta$ as given in Eqn.~\ref{eq:IntervalDgm} are the induced maps on the connected components between the noted intervals.}
  \label{fig:GraphsWithIntervals}
\end{figure}

\begin{proof}
We will explicitly construct the map $\Phi$, and show it is a bijection.
Given a vertex $v$ with $f(v) = a$, this is done by finding a vertex in $S_\e(X)$ with either function value $a + \e$ or $a-\e$ depending on the type of $v$, and showing that this matching between the vertex sets is unique.

First, assume $v$ is a local minimum.
Building the diagram of Eq.~\ref{eq:IntervalDgm} with $b = a-\e$ with corresponding maps $\alpha$ and $\beta$, we see that the connected component of $v$ in $\pi_0f\inv[b]^{\e+\delta}$ is not in the image of $\alpha$.
Thus, there is a vertex $w$ in $S_\e(X,f)$ with $f_\e(w) = b$, and we define $\Phi(v) = w$.
Further, the above construction gives us that the down-degree of $w$ is zero.
Noting that $\delta$ was chosen sufficiently small in Eq.~\ref{eq:IntervalDgm} so that  there is no additional vertex between $b+\e$ and $b+ \e+\delta$, we also have that $\beta$ is an isomorphism.  Therefore, the up-degree of $w$ is 1, implying that $w$ is itself a local minimum vertex.

We can use the symmetric argument with $b = a+\e$ in the case where $v$ is a local maximum to find that its connected component is not in the image of $\beta$, and thus there is a local maximum vertex $w$ at height $a+\e$ in $S_\e(X,f)$.
We again define $\Phi(v) = w$

Next, assume that $v$ is a down fork which is not part of a $\rel_1$ pair with lifetime less than $2\e$; i.e.~$v \not \in W$.
In this case, we start by building Eqn.~\ref{eq:IntervalDgm} with $b= a-\e$.
Consider the two lower edges of $v$, and the points $x$ and $y$ on these edges at height $a-\delta$.
We first show that $x$ and $y$ are in different connected components of $\pi_0 f\inv ([b-\delta]^\e)$.
If they are in the same connected component, there is a path in $f\inv ([b-\delta]^\e)$ connecting the two points, let $u$ be the vertex with lowest function value on this path.
By choice of $\delta$, we know that there is no vertex in the interval $[b-\e -\delta, b-\e)$, so the height difference between $v$ and $u$ is at most $2\e$.
However, in this case, extended persistence would pair the vertex $u$ (the lowest possible vertex) with $v$ as a $\rel_1$ pair which has lifetime at most $2\e$.
This means $v \in W$, contradicting our assumption.

As we now know that $x$ and $y$ are in two different connected components of $f\inv([b-\e]^\delta)$, there are two elements of $\pi_0f\inv([b-\e]^\delta)$ whose image under $\alpha$ is the component containing $v$.
Because there are no vertices of $(X,f)$ in  $f\inv((b+\e,b+\delta+\e])$ by our choice of $\delta$, there is a single component in $\pi_0f\inv([b+\e]^\delta)$ mapping to the component of $v$ under $\beta$.
Thus, we  have a vertex $w$ in $S_\e(X,f)$ at height $b$ with down degree 2 and up degree 1.
We set $\Phi(v) = w$.
See Fig.~\ref{fig:GraphsWithIntervals} for an example.

By symmetry, we can use this same argument with $b=a+\e$ for an up fork vertex, so that for any up fork $v$ in $(X,f)$, there is an up fork $w$ in $S_\e(X,f)$ at height $a + \e$, and set $\Phi(v) = w$.

Now that $\Phi$ has been defined for all required vertices, show that $\Phi: V \setminus W \to V(S_\e(X))$ is a bijection.
For surjectivity, let $u$ be a vertex in $S_\e(X)$ at height $b$.
By assumption on our Reeb graphs, this must have at least one of the up or down degree 0 or 2 (i.e., not both degree 1).
Assume first that the down degree is 0 and again setup the diagram of Eqn.~\ref{eq:IntervalDgm}.
Because of vertex $u$, we must have that there is a connected component $[v] \in \pi_0 f\inv([b]^\e)$ which is not in the image of $\alpha$; but this is exactly the requirement for having a local minimum vertex at height $b+\e$ in $(X,f)$, and so $\Phi(v) = u$.
A symmetric argument can be made for local maxima.

Similarly, assume $u$ is a vertex in $S_\e(G)$ at height $b$ which has down degree 2.
Then there are two connected components in $f\inv[b-\delta]^\e$ whose image under $\alpha$ is the component represented by $u$, and thus there is a vertex in $(X,f)$ at height $b+\e$ with $\Phi(v) = u$.
Again, a symmetric argument can be used in the case of up-degree 2; thus $\Phi$ is surjective.

For injectivity, we note that by assumption, our Reeb graph is generic, meaning that we cannot have two vertices at the same height.
If there was a vertex $u \in S_\e(X,f)$ at height $b$ and two vertices $v,w \in (X,f)$ with $\Phi(v) = \Phi(w) = u$, this would imply that one must be at height $b+\e$ and the other at height $b-\e$; without loss of generality assume these are $v$ and $w$ respectively.
But then our value $\e = 2|a_i - a_j|$, where $f(v) = a_i$ and $f(w) = a_j$ are critical values, contradicting our assumptions on $\e$.
Thus $\Phi$ is a bijection, and $S_\e(X,f)$ is generic.
\end{proof}

Our next proposition proves that $\Phi$ does not affect the persistence pairing of critical points in the Reeb graph.

\begin{theorem}
\label{thm:pairpoints}
Given a generic Reeb graph, for vertices $u,v \in V(X,f)\setminus W$  which are paired under extended persistence, $\Phi(u)$ and $\Phi(v)$ are paired and of the same type in the extended persistence diagram of $S_\e(X,f)$.
\end{theorem}

\begin{proof}
We break this into cases based on the four types of paired points in the extended persistence diagram.
In each case, we assume we have paired vertices of $(X,f)$, $u$ and $v$.
For notation, we assume $f(u) = a_i$, $f(v) = a_j$, and $a_i \leq a_j$.

From prior work describing extended persistence of Reeb graphs~\cite{Agarwal2006,Cohen-Steiner2009a,Cole-McLaughlin2003} we have a simple pairing for all the vertices.
Namely, the global maximum in any component is paired with its corresponding global minimum ($\Ext_0$ points).
A down fork is paired with the highest up fork that spans a loop with it in the graph, if one exists ($\Ext_1$ points).
Any remaining downfork is paired with the higher of the two minima for the two components in its sublevel set ($\Ord_0$ points);
similarly any remaining upfork is paired with the lower of the two maxima of the two components of its superlevel set ($\Rel_1$ points).

If the pair is in $Ext_0$, we have a point $(a_i,a_j)$ in the diagram.
Further, $u$ and $v$ are the global  minimum and maximum respectively of a connected component of $X$.
From Theorem~\ref{thm:bijection}, we know that $S_\e(X,f)$ has vertices $\Phi(u)$ and $\Phi(v)$ at height  $a_i-\e$ and $a_j+\e$.
Moreover, any other critical point $b$ in this component of the original graph has value $a_i < b < a_j$, so any critical point in the related component of  $S_\e(X,f)$ is  between $a_i - \e$ and $a_j + \e$.
This implies $\Phi(u)$ and $\Phi(v)$ are the global maximum and minimum of a component of the smoothed Reeb graph, and hence they are paired by extended persistence in $\Ext_0$.

If the pair is in $\Ord_0$, we have a point $(a_i,a_j)$ in the diagram, where $u$ is a local minimum, and $v$ is an ordinary down fork.
Let $C_1$ and $C_2$ be the connected components of $f \inv (\infty, a_j)$ below $v$.
Let $u$ and $u'$ be the minimum function value vertices of these two connected components respectively.
Since $v$ is paired with $u$, this implies that $f(u') < f(u)$.
After smoothing $(X,f)$  by $\e$, by Theorem~\ref{thm:bijection} we know that $S_\e(X,f)$ has local minimum $\Phi(u)$ at the critical value $a_i - \e$, local minimum $\Phi(u')$ at $f(u')-\e$, and downfork $\Phi(v)$ at  $a_j - \e$.
Consider the connected component(s) below $\Phi(v)$ in $\tilde f\inv(-\infty, a_j-\e)$.
Note that by construction, the Reeb quotient map of the thickened space $X \times [-\e,\e]$ maintains connected components, so
$$
\pi_0[q]:
\pi_0(q \inv (\tilde f \inv(-\infty, a_j-\e)
\cong
\pi_0\tilde f \inv(-\infty, a_j-\e)
$$
gives an isomorphism.

The sets $C_1 \times \{-\e\}$ and $C_2\times \{-\e\}$ must be disconnected in $q \inv (\tilde f \inv(-\infty, a_j-\e) \subset X \times [-\e,\e]$ because $C_1$ and $C_2$ are disconnected in $X$.
Thus, by the isomorphism $\pi_0[q]$, they must map to different connected components below $\Phi(v)$.
First, this implies that $\Phi(v)$ must be an essential downfork.
Second, we must have $\Phi(u)$ and $\Phi(u')$ in these connected components, and they must be the minimum function value vertices of each of the connected components.
We know that $f(\Phi(u))\geq f(\Phi(u'))$, so this implies that $\Phi(u)$ is paired with $\Phi(v)$.

The argument for $\Rel_1$ is the same as that of $\Ord_0$, with super- and sublevel sets switched.
Thus our final case is when $u$ and $v$ are paired in $\Ext_1$.
We have already shown that all points in $\Rel_1, \Ord_0$, and $\Ext_0$ from $X$ stay paired under $\Phi$, so all that remains is to be sure $\Phi(u)$ and $\Phi(v)$ cannot pair with any other points from $\Ext_1$.
Since $u$ is the highest up fork that spans a loop with $v$, there are exactly two connected components in $f^{-1}((a_i,a_j))$ which attach to $u$ and $v$ in $X$ via the inclusion maps into $f^{-1}([a_i,a_j])$.
Therefore, by Lemma~\ref{lem:homotopy_equiv}, we have the commutative diagram
\begin{equation*}
\begin{tikzcd}
f_\e\inv(a+\e,b-\e) \ar[r,"\simeq"] \ar[d, hook]& f\inv(a,b) \ar[d, hook]\\
f_\e\inv[a+\e,b-\e] \ar[r,"\simeq"]& f\inv[a,b]
\end{tikzcd}
\end{equation*}
and thus there are exactly two connected components in $f_\e^{-1}((a_i+\e,a_j-\e))$ which attach to $\Phi(u)$ and $\Phi(v)$ in $S_\e(X,f)$ via the inclusion maps.
Therefore, $\Phi(u)$ and $\Phi(v)$ will remain paired in $\Ext_1$.
\end{proof}

Again, the prior result is only proven here for generic graphs, but the proof can be adapted to work in non-generic graphs as well.  In that case, a vertex of the graph simply corresponds to multiple critical values, so that any vertex of degree $d$ in the graph will appear in $d-2$ persistence pairs in the diagram; the case analysis correspondingly becomes more complex.

Finally, we arrive at the true main result of this section, where we can use the characterization of the movement of the  critical points to keep track of movement in the persistence diagram.

\begin{figure}
  \includegraphics[width = .6\textwidth]{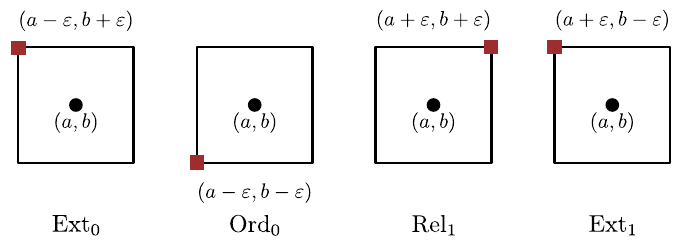}
    \centering
  \caption{A point $(a,b)$ in the diagram for the Reeb graph $(X,f)$ moves to a point in the diagram of $S_\e(X,f)$ depending on which type of persistence point it represents.}
  \label{fig:smootheddiagram}
\end{figure}

\begin{corollary}
 \label{cor:PointsMove}
Consider a generic Reeb graph $(X,f)$, a value $\e$, and the bijection $\Phi$ as in Theorem~\ref{thm:bijection}.
For every point $(a,b) = (f(v),f(u))$ in the persistence diagram of $(X,f)$, the corresponding point $(f(\Phi(v)),f(\Phi(u)))$  in the persistence diagram of $S_\e(X,f)$ located (depending on its type) as follows:
\begin{center}
    \begin{tabular}{c|l}
         $Ext_0$& $(a- \e, b+\e)$ \\
         $Ord_0$& $(a- \e, b-\e)$ \\
         $Rel_1$& $(a+ \e, b+\e)$ \\
         $Ext_1$ &
            $
            \begin{cases}
            (a+\e,b-\e) & \text{if } \frac{(a+\e,b-\e)}{2}< \e\\
            \text{removed} & \text{if }  \frac{(a+\e,b-\e)}{2}> \e  .
            \end{cases}
            $
    \end{tabular}
\end{center}

\end{corollary}

See Fig.~\ref{fig:smootheddiagram} for a visual representation of the movement of the points; and Fig.~\ref{fig:smoothed23} for the movements of point in a diagram where smoothing without truncation is represented by the case where $\tau = 0$.

\begin{proof}
This proof is bookkeeping to keep track of the types of top and bottom points forming each pair in the persistence diagram and noting how they move via Theorem~\ref{thm:bijection} and Theorem~\ref{thm:pairpoints}.
\end{proof}

\subsection{Truncating Smoothed Reeb Graphs}

We next consider truncated smoothing, and prove an analogous characterization of its impact on the persistence diagram.
We know that $\beta_0(X,f) = \beta_0(S_\e^\tau(X,f))$, since their $\pi_0$ groups are isomorphic~\cite{chambers2020family}.
We begin by proving an analogous result on $\pi_1(S_\e^\tau(X,f))$.

\begin{lemma} \label{lem:pi1}
Given $(X,f)$ and $0\leq\tau < 2\e$
then $\pi_1[\eta]$ induces an isomorphism $\pi_1(S_\e^\tau(X,f)) \cong \pi_1(S_\e(X,f))$.
\end{lemma}

\begin{proof}

Consider a loop in $(X,f)$ with critical values $a_i$ and $a_j$.
By Theorem~\ref{thm:bijection},  the loop either disappears under the smoothing functor (if its height is less than $\e/2$) or is still present in the smoothed graph (under the bijection $\Phi$).
By Prop.~4.3 and Lemma 4.4 of \cite{chambers2020family}, for $\tau \le 2\e$, truncation of the smoothed graph will not reach any essential fork.
Hence each loop maps to a unique loop in truncated graph, under the natural inclusion from $S^\tau_\e(X,f) \hookrightarrow S_\e(X,f)$, and the resulting isomorphism follows since no new loops can be created under truncation.
\end{proof}

\begin{figure}
\centering
  \includegraphics[width = .6\textwidth]{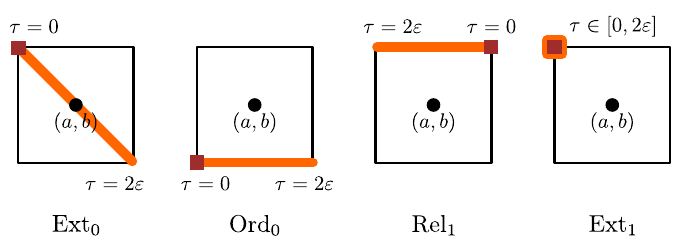}
  \caption{A point $(a,b)$ in the diagram for the Reeb graph $(X,f)$ moves to a point in the diagram of $S_\e^\tau(X,f)$ depending on which type of persistence point it represents. The red dot again indicates the behavior of smoothing (when $\tau = 0$), and the orange lines indicate the range of possible values in the diagram reachable by different values of $\tau < 2\e$.}
  \label{fig:smoothedTruncdiagram}
\end{figure}

The goal of the following proposition is to see how truncation affects  the four types of persistence points ($\Ext_0$, $\Ord_0$, $\Rel_1$ and $\Ext_1$) after smoothing, with assumptions to ensure that truncation does not change the topology of a given graph.  See Fig.~\ref{fig:smoothedTruncdiagram} for a visualization of the theorem, which constrains the
effect of  truncated smoothing on the different points of the persistence diagram.

\begin{proposition}
\label{prop:truncatedPointsMove}
Fix $0 \leq \tau \leq 2\e$.
Suppose $(X,f)$ is a generic Reeb graph.
For every point $(a,b)$ in the persistence diagram of the  Reeb graph,  the corresponding point in the persistence diagram of the truncated smoothed Reeb graph $S_{\e} ^{\tau}(X,f)$ is
\begin{center}
    \begin{tabular}{c|l}
         $Ext_0$& $(a-\e+ \tau, b+\e-\tau)$ \\
         $Ord_0$& $(a-\e+ \tau, b-\e)$ \\
         $Rel_1$& $(a+ \e, b+\e-\tau)$ \\
        $Ext_1$ & $(a+\e,b-\e)$
    \end{tabular}
\end{center}
if the new point is on the same side of the diagonal as $(a,b)$, and the point is removed completely if not. %
\end{proposition}

\begin{proof}

By Lemma~\ref{lem:pi1} and our assumptions on $X$, we know that truncation on the smoothed graph moves any critical point that a local minimum up by $\tau$, any local maximum down by $\tau$, and leaves up- and down-forks unchanged.  The proof then follows from the same bookkeeping as in Corollary~\ref{cor:PointsMove}, after tracking these new critical values.
\end{proof}

\begin{figure}
\includegraphics[width = \textwidth]{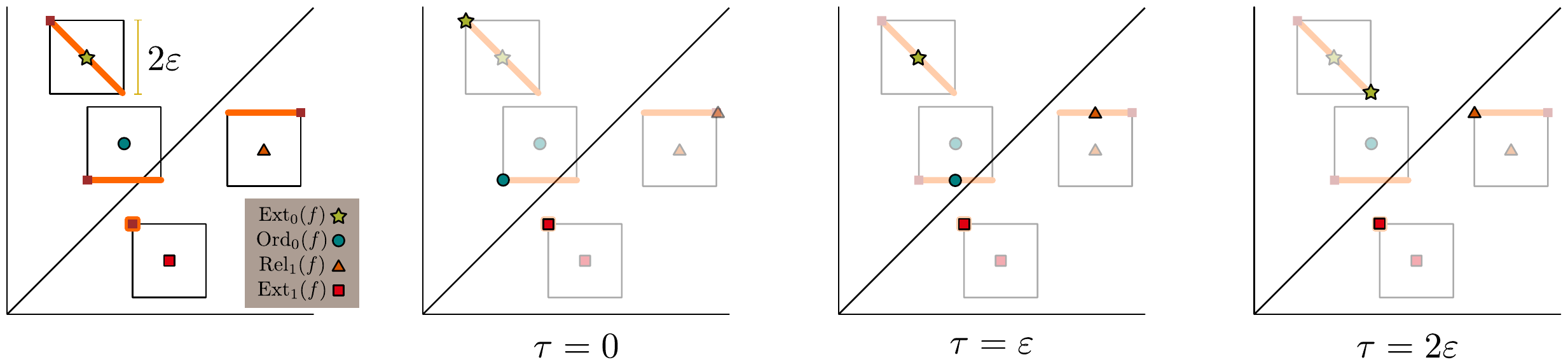}
    \centering
  \caption{Given an initial extended persistence diagram (left), the three diagrams at right are the resulting diagrams after smoothing by $\e$ and truncating by $\tau = 0, \e, 2\e$ respectively. Note that in this example, for $\tau = 2\e$ the point in $\Ord_0$ is removed completely since its location on the box would be on the other side of the diagonal. }
  \label{fig:smoothed23}
\end{figure}

See Fig.~\ref{fig:smoothedTruncdiagram} for a visualization of the behavior of different types of point sin the diagram.
We refer to Fig.~\ref{fig:smoothed23} for further visualization of how this looks on a full diagram.

\section{Inverse problems on Reeb graphs}
\label{sec:inverseproblems}

We are now in a position to make use of this characterization to provide a solution to the time varying inverse problem in a particular restricted setting.
In the most general setting, our problem can be stated as follows.
We assume we are given a path in persistence diagram space, colloquially known as a vineyard, which is data $\gamma: \R_{\geq0} \to \Pers$.
We assume this path is a continuous map under the bottleneck distance, and that we have also been provided with an initial Reeb graph $R_0 = (G_0,f_0)$.
The goal is to find a path in Reeb graph space, continuous with respect to the interleaving distance \cite{deSilva2016}, $\Gamma:\R_{\geq0} \to \Reeb$, for which $\Gamma(0) = R_0$ and $\PP(\Gamma(t)) = \gamma(t)$.

We  restrict our view to essentially creating a notion of piecewise linear paths, defined by linear interpolation between the path defined at discrete times $0 = t_0 < t_1 < t_2 < t_3 \cdots$.
For the sake of notation, we will denote $\gamma(t_i) = D_i$, and for the Reeb graph path we will construct, $\Gamma(t_i) = R_i = (G_i,f_i)$.
Geodesics in persistence diagram space are defined by matchings arising from the bottleneck distance computation.
While a minimum cost matching is not unique in this setting, say we have a matching $M_i: D_i \to D_{i+1}$, then the geodesic between diagrams $D_i $ and $D_{i+1}$ is given by sliding each point $x$ at constant speed along the line between $x \in D_i$ and $M_i(x) \in D_{i+1}$.

Following Prop.~\ref{prop:truncatedPointsMove}, we define the map $\Psi_\e^\tau: \Pers \rightarrow \Pers$ on diagrams by defining a map on each point $(a,b) \in D$ as follows:

\begin{equation*}
\phi_\e^\tau(a,b) =
    \begin{cases}
      (a - \e +\tau, b+ \e -\tau) & \text{if the type of }(a,b) \text{ is } Ext_0 \text{ in } D \\

   (a- \e + \tau, b-\e) & \text{if the type of }(a,b) \text{ is } Ord_0 \text{ in }  D\\
    (a + \e, b+ \e - \tau)  & \text{if the type of }(a,b) \text{ is } Rel_1 \text{ in } D\\
   (a + \e, b- \e) & \text{if the type of }(a,b) \text{ is } Ext_1 \text{ in } D.
    \end{cases}
    \label{mapcases}
\end{equation*}
Then the map $\Psi^\tau_\e$ is defined by
$$\Psi^\tau_\e(D) = \{ \phi^\tau_\e(x,y) \mid (x,y) \in D \text{ if }\phi^\tau_\e(x,y)\text{ is on the same side of the diagonal }\Delta\text{ as }(x,y)  \}.$$

We can abuse notation and also view this construction as a matching on the points of related diagrams themselves.
Given an $\e$ and $\tau$, all points of the same type ($\Ord_0$, $\Rel_1$, etc) move in the same direction.
Thus, for a given direction vector $\Vec{v}$, which depends on $\e$, $\tau$, and point type, we can define an updated single-type diagram $D_{\Vec{v}}$ by:
\begin{equation*}
    D_{\Vec{v}}=D+ {\Vec{v}} := \{x  +\Vec{v} \mid x \in D \text{ with } x\text{ and }x + \Vec{v} \text{ on the same side of the diagonal}  \}.
\end{equation*}
Given this updated diagram, we have a matching defined by:
\begin{equation}
\label{eqn:phi}
    \omega: x \mapsto:
    \begin{cases}
    x +\Vec{v}  &  x \in D_{\Vec{v}} \\
    \Delta   &   \text {otherwise} .
    \end{cases}
\end{equation}
With this setup, we begin with the following theorem, which proves directly that our truncated smoothing map in fact is at least locally  bottleneck optimal, motivating our use of this operation for inverse problems.

\begin{theorem}
\label{thm:bottleneckbound}
For $\Vec{v}$ with magnitude
$$
|\Vec{v}| \leq
 \frac{1}{2}\min_{x \in D} \left\{
    \lVert x-\Delta \rVert_\infty,
    \min_{y \in D} \lVert x-y \rVert_\infty
\right\},
$$
and a diagram $D$ with only points of a single type,
the matching induced by $\omega$ achieves the bottleneck distance between $D$ and $D_{\Vec{v}}$.
\end{theorem}

\begin{proof}
Note that by assumption, no points of $D$ have been removed in $D_{\Vec{v}}$, since the magnitude of $\Vec{v}$ cannot reach the diagonal.
As a result, the  matching $\omega$ is a bijection between only the off-diagonal points in the diagrams  sending $a$ to $a+\Vec{v}$.

We will prove this theorem by contradiction.
Assume that $\omega$ is not a bottleneck matching; in particular, this means that $d_B(D,D_{\Vec{v}})< \|\Vec {v}\|_\infty$.
So there exists another matching  $\psi: D \to D_{\Vec{v}}$ that achieves bottleneck distance.
Specifically, for all $a \in D$, $\|a-\psi(a)\|<\|\Vec{v}\|_\infty$.
If any point $a$ matches to the diagonal $\Delta$ under $\psi$, then by assumption on $\|\Vec{v}\|_\infty$, we have that  $\|a-\psi(a)\| = \lVert a-\Delta \rVert \geq \|\Vec{v}\|_\infty$, contradicting our assumption.
So we may assume every point will match to an off-diagonal point under $\psi$.

\begin{figure}
    \centering
    \includegraphics[width = .3\textwidth]{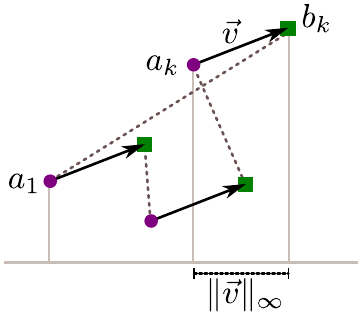}
    \caption{An example of the two matchings $\omega$ (shown in solid arrows) and $\psi$ (shown as dotted lines) from case (R) of the proof of Thm.~\ref{thm:bottleneckbound}.
    Their projections down to the $x$-axis are shown, and the direction of $\Vec{v} \in [-\frac{\pi}{4},\frac{\pi}{4}]$.}
    \label{fig:MatchingCycle}
\end{figure}

We can think of the matchings $\omega$ and $\psi$ as being defined as edge sets $M_\omega$ and $M_\psi$ in a bipartite graph with vertex sets given by the off-diagonal points of $D$ and $D+\Vec{v}$, respectively.
Then we can consider the symmetric difference $M_\omega \Delta M_\psi$ of the matchings as the set of pairs $(a,b)$ for which $b = \omega(a)$ or $b = \psi(a)$ but not both.
By \cite[Lem.~3.1.9]{West2001}, every component of the symmetric difference of two (not necessarily perfect) matchings is a path or an even cycle.
Because $\psi$ and $\omega$ are perfect matchings, every vertex in $M_\omega \Delta M_\psi$ has degree 0 or degree 2; hence, each component must be an even cycle, since a path requires degree 1 vertices.

Choose one of these cycles $C$.
We will find a pair $(a,\psi(a))$ in this cycle $C$ for which $\|a-\psi(a)\|_\infty>\|\Vec{v}\|_\infty$, thus showing the required contradiction.
The choice of $a$ will be dependent on the direction of $\Vec{v}$, however, so we split our work into four cases depending on the angle of $\Vec{v}$:
(R) $[-\frac{\pi}{4},\frac{\pi}{4}]$;
(U) $[\frac{\pi}{4},\frac{3\pi}{4}]$;
(L) $[\frac{3\pi}{4},\frac{5\pi}{4}]$; and
(D) $[\frac{5\pi}{4},-\frac{\pi}{4}]$.

Assume case (R), so $\Vec{v}$ is essentially going to the right.
In this case, the change in the $x$-axis determines the $L_\infty$ bottleneck matching distance for $\omega$.
Let $a_1$ be the point in cycle $C$ with smallest $x$-coordinate.
Define $b_i = \omega(a_i)$, and enumerate the cycle $C$ starting from $a_1$, so
$C = a_1b_1a_2b_2\cdots a_kb_k$ with $\psi(a_i) = b_{i-1}$ for $i>1$ and $\psi(a_1) = b_k$.
Because we chose $a_1$ for its $x$-coordinate and by assumption on the direction of $\Vec{v}$, we know that $(a_1)_x \leq (a_k)_x < (b_k)_x$ where $\bullet_x$ denotes $x$-coordinate.
Then in this case, we have
\begin{equation*}
\|a_1-\psi(a_1)\|
    = (b_k)_x - (a_1)_x
    \geq ((b_k)_x - (a_k)_x) + ((a_k)_x-(a_1)_x)
    \geq \|\Vec{v}\|_\infty
\end{equation*}
giving us the desired contradiction.
See~Fig.~\ref{fig:MatchingCycle} for an example.

The remaining cases follow the same structure as (R) with mild tweaks.
For case (L), we set $a_1$ to be the point with largest $x$-coordinate.
For the cases (U) and (D), we choose $a_1$ to be the point with smallest and largest $y$-value, respectively.
\end{proof}

Next, we turn our attention to when a sequence of diagrams (i.e. a \emph{vineyard}) is realizable by Reeb graphs using smoothing and truncation operations.

\begin{definition}
A sequence of input diagrams $\{ D_i\}_{i=0}^N$ is \emph{admissible} if for every $i$, there exists a pair $(\e_i, \tau_i)$ with $\tau_i < 2\e_i$, such that $D_{i+1} = \Psi^{\tau_i}_{\e_i}(D_i)$.
\end{definition}

\begin{figure}
    \centering
    \includegraphics[width = \textwidth]{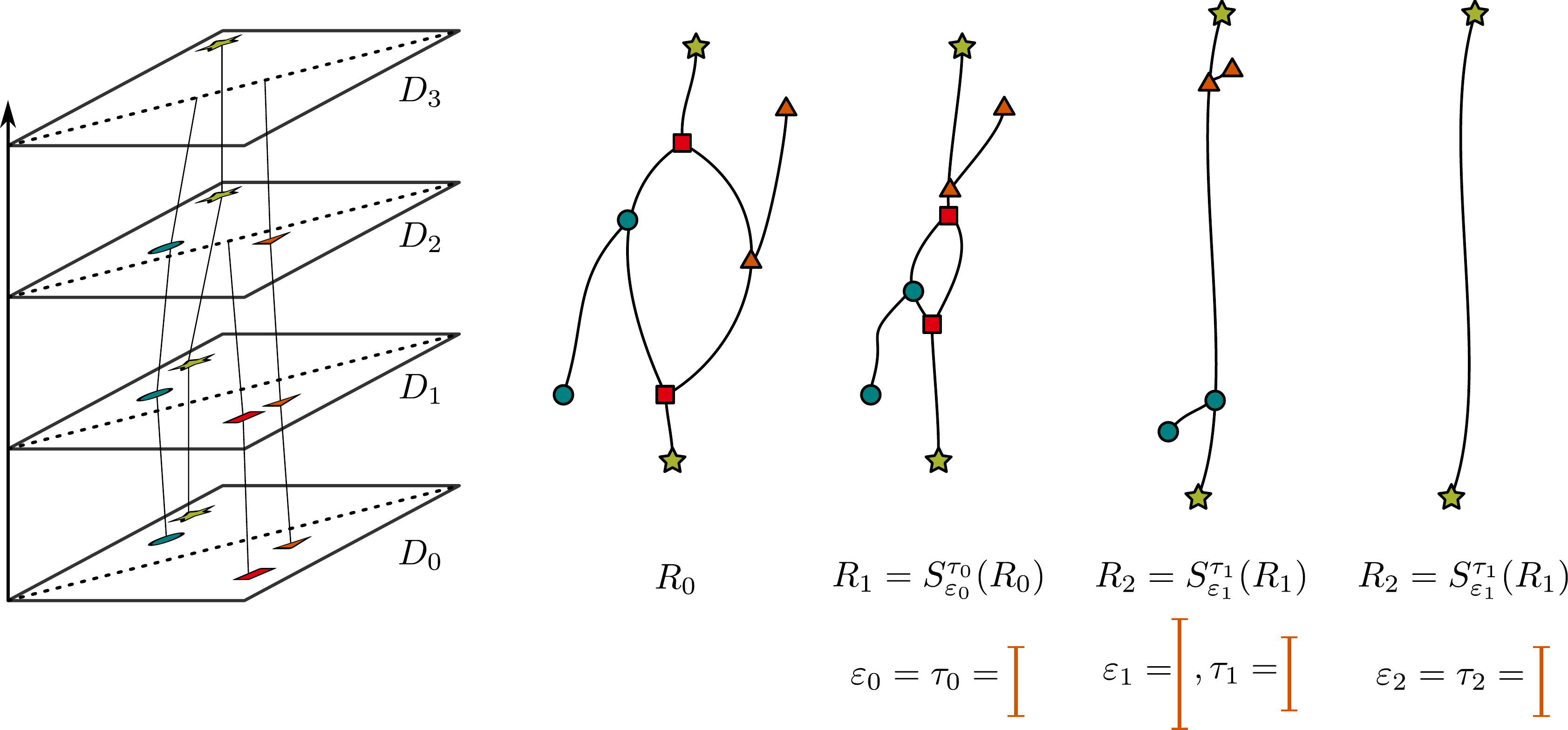}
    \caption{An example of a sequence of admissible diagrams (left) realized by a series of Reeb graphs with $\e_i$ and $\tau_i$ shown below. }
    \label{fig:vineyard}
\end{figure}

For example, see Figure~\ref{fig:vineyard}, where pairs of admissible diagrams are shown for different $\e,\tau$ parameters.
This input restriction is built to use Proposition~\ref{prop:truncatedPointsMove} to provide a solution to the inverse problem.

\begin{theorem}\label{thm:sequence}
Assume $\{ D_i\}_{i=0}^N$ is admissible and Reeb graph $R_0$ is given so that $\PP(R_0) = D_0$.
Then there is a sequence of Reeb graphs $R_i$ for which $\PP(R_i) = D_i$.

\end{theorem}

\begin{proof}
The proof proceeds by induction.
For $i=0$, and since the $D_i$'s are admissible, by Proposition~\ref{prop:truncatedPointsMove} we have a pair $(\e_0,\tau_0)$ such that
$D_1$ is the diagram of $S_{\e_0}^{\tau_0}(R_0)$~i.e $\Psi_{\e_0}^{\tau_0}(D_0)= D_1$.
We can then  define the first Reeb graph to be $R_1 = S_{\e_0}^{\tau_0}(R_0)$, which has the property that $\PP(R_1) = D_1$ by constructing the truncated smoothing.

Next we for any $i\geq 0$, we assume we have a Reeb graph $R_i$ such that $\PP(R_i) = D_i$, and wish to build $R_{i+1}$ such that $\PP(R_{i+1})= D_{i+1}$.
We proceed in the same manner as the base case by finding $(\e_i,\tau_i)$ using the fact that the $D_i$ collection is admissible, and setting $R_{i+1} = S_{\e_i}^{\tau_i}(R_i)$.
Then by Proposition~\ref{prop:truncatedPointsMove} $\PP(R_{i+1}) = \Psi_{\e_i}^{\tau_i}(D_i) = D_{i+1}$.
\end{proof}

\begin{corollary}
Given an admissible sequence of diagrams $\{D_i\}_{i=0}^N$, we can extend this to a piecewise linear vineyard given by
$\gamma(i+t) = D_i+t \Vec{v_i}$ for integer $i$ and $t \in [0,1)$, where $\Vec{v_i}$ is set for the pair $D_i,D_{i+1}$ by the formula in Equation~\ref{eqn:phi} and is distinct by diagram subtype.
Then the path of Reeb graphs
$$\Gamma(i+t) = S_{t\e_i}^{t\tau_i}(R_i)$$
realizes this path; i.e.~$\mathcal{P}(\Gamma(i+t)) = \gamma(i+t)$.

\end{corollary}

\begin{proof}
This corollary follows from Theorem~\ref{thm:sequence} as follows.  First, note that if the pair $D_i$, $D_{i+1}$ is admissible when using $\e_i$ and $\tau_i$, we get the direction $\Vec{v_i}$ for each type of point as in Equation~\ref{eqn:phi}.  Then, we can linearly interpolate along $\Vec{v}_i$ in the diagram to get an intermediate diagram, and Theorems~\ref{thm:bijection} and~\ref{thm:pairpoints} tell us there are intermediate values of $\e$ and $\tau$ that realize the smoothing and diagram for any point along the interpolation.
\end{proof}

\section{Conclusions and future directions}
\label{sec:conclusion}

In this paper, we have provided a complete characterization of the behavior of vertices in a Reeb graph under the truncation and smoothing operations with respect to their extended persistence.
We showed how this classification can be used to open doors to further utilization of the smoothing procedure itself.
In particular, as we have seen, this characterization can be translated into simple descriptions of the available paths in extended persistence diagram space under these transformations,
yielding tractable solutions to (an admittedly rather restricted version of) inverse problems in topological data analysis.

This work suggests many possible  future directions to study.
While we have only dealt with piecewise linear vineyards in this work, it seems likely that there will be ways to loosen our restrictions in Theorem~\ref{thm:sequence}.
Our analysis of smoothing suggests many possible relaxations, where for example different sections of the Reeb graphs are smoothing and truncated (or even ``untruncated" and ``unsmoothed") by varying amounts or where more relaxed constraints on admissible could yield approximate solutions to the inverse problem.
This broader framework of piecewise linear paths would be interesting if it can be used to approximate more general vineyards, yielding the potential for approximate solutions of a more general inverse problem.
In addition, while we have chosen to analyze using the bottleneck distance here in Theorem~\ref{thm:bottleneckbound}, there are many other related metrics on both Reeb graphs and persistence diagrams to consider, and we conjecture that the paths generated by truncated smoothing are locally  optimal under other metrics.

\paragraph{Acknowledgements:} This work was partially supported by NSF grants CCF-1907612, CCF-1614562, DBI-1759807, DMS-1800446, CMMI-1800466, and CCF-1907591.

\printbibliography
\end{document}